# Current-controlled creations, deletions, and topological transformations of a single magnetic antiskyrmion in nanostructured cells

Yaodong Wu[1], Jialiang Jiang[2,3*], Lingyao Kong[2], Wei Liu[4], Huanhuan Zhang[2], Shouguo Wang[5], Mingliang Tian[2,3], Haifeng Du[3*], and Jin Tang[2,3*]

[1]School of Physics and Materials Engineering, Hefei Normal University, Hefei 230601, China

[2]State Key Laboratory of Opto-Electronic Information Acquisition and Protection Technology, School of Physics, Anhui University, Hefei, 230601, China

[3]Anhui Provincial Key Laboratory of Low-Energy Quantum Materials and Devices, High Magnetic Field Laboratory, HFIPS, Chinese Academy of Sciences, Hefei 230031, China

[4]Information Materials and Intelligent Sensing Laboratory of Anhui Province, Institutes of Physical Science and Information Technology, Anhui University, Hefei, 230601, China

[5]Anhui Provincial Key Laboratory of Magnetic Functional Materials and Devices, School of Materials Science and Engineering, Anhui University, Hefei 230601, China

*email: jjl2024@ahu.edu.cn; duhf@hmfl.ac.cn; jintang@ahu.edu.cn;

**Abstract**

Topological magnetic solitons have emerged as promising candidates for information carriers in spintronic devices, thanks to their fascinating electromagnetic properties. For fundamental device applications, the ability to electrically manipulate individual solitons is crucial. However, electrical manipulation of single antiskyrmions has been rarely demonstrated. In this work, we present current-controlled manipulations, encompassing the creation, deletion, and topological transformation of a single antiskyrmion within FeNiPdP nanostructured cells at room temperature. This nanostructure is uniquely designed with dimensions of about 400 nm in width and length, enabling the stabilization of a single antiskyrmion. By simply adjusting the density of nanosecond single-pulsed currents, we achieve the reversible creation and deletion of single antiskyrmions. Moreover, we uncover a rich variety of current-controlled topological transformations among individual antiskyrmions, skyrmions, bubbles, and ferromagnetic states. Our experimental findings are corroborated by micromagnetic simulations, highlighting the pivotal role of current-induced combined effects, such as spin transfer torque and Joule heating. Our results hold potential for advancing antiskyrmion-based device applications.

Topological magnetic solitons, embodied by skyrmions and antiskyrmions, are localized particle-like magnetic structures that boast small size, fascinating electromagnetic properties, and tunability through external electrical stimuli[1-5]. These magnetic solitons have the potential to serve as versatile information carriers for various functions, including memory, storage, and logic operations[6-9].

The electrical manipulation of skyrmions and antiskyrmions has been a long-standing area of interest for practical device applications. Recent research has demonstrated diverse electrical control over skyrmions, encompassing their creation[10-17], deletion[10-14], and movement[14-22], which align with the fundamental device functions of writing, deleting, and addressing, respectively. Vortex-like magnetic solitons are typically described by the polarity $P$, vorticity $v$, and topological charge $Q$, and $Q = Pv$. Magnetic skyrmions and antiskyrmions are distinguished by their vorticity, which is consistent across other condensed matter systems.[23-25] Magnetic skyrmions are fundamentally categorized by their helicity into two types: Bloch and Néel. In contrast, magnetic antiskyrmions possessing an opposite sign of vorticity, observed in noncentrosymmetric magnets with anisotropic $D_{2d}$ interactions, possess opposite topological charges $Q$ for the same polarity $P$.[26-28] We note that the classification of skyrmions and antiskyrmions in FeGe magnets, as defined in prior work,[29] is based on topological charge—a scheme that differs fundamentally from the one employed in our present study.

Antiskyrmions have been observed in a handful of non-centrosymmetric magnets, like Mn-Pt-Sn and Fe-Ni-Pd-P alloys[30-32]. Due to intricate competing magnetic interactions, it has been established that multiple magnetic solitons, including dipolar skyrmions and bubbles, can coexist with antiskyrmions[33-37]. It is important to note that skyrmion stabilization in antiskyrmion-hosting material systems arises not from the Dzyaloshinskii-Moriya interaction (DMI), but instead shares the same origin as traditional bubbles in centrosymmetric uniaxial magnets. Magnetic bubbles are typically observed in centrosymmetric uniaxial magnets.[38,39] They fall into two categories: type-I bubbles and type-II bubbles. Because type-I bubbles share identical polarity, vorticity, and integer topological charge ($Q$ = 1 or −1) with chiral-Bloch-twisted skyrmions, they are typically classified as 'skyrmions' in antiskyrmion-hosting material systems. Conversely, topologically trivial type-II bubbles ($Q$ = 0) retain the conventional 'bubble' designation.[33-37,40,41] Moreover, the additional anisotropic DMI in $D_{2d}/S_4$-symmetric systems distorts type-I and II bubbles from circular to elliptical shapes. Different from the dipolar skyrmions and bubbles, magnetic antiskyrmions with square shapes are mainly supported by DMI.

A pioneering work has demonstrated that coexisting magnetic solitons enable information encoding via distinct topological charges $Q$.[42] Consequently, manipulating these solitons topologically represents a promising approach, previously explored using external magnetic fields or thermal gradients,[33-36,43] these methods lack direct compatibility with high-speed electronic devices. Theoretical studies have revealed intriguing current-induced dynamics of antiskyrmions, such as the antiskyrmion Hall effect and the creation of skyrmion-antiskyrmion pairs.[42-47] Beyond electrical creation and deletion of antiskyrmions, these systems also enable electrically driven skyrmion-antiskyrmion and antiskyrmion-bubble transformations, a capability not previously achieved in conventional skyrmion-hosting materials. The experimental electrical manipulation of antiskyrmions remains challenging, with only recent studies exploring the current-induced motion of antiskyrmions confined within a helix[48,49]. The controlled electrical creation, deletion, and topological transformations of individual magnetic antiskyrmions have yet to be achieved.

In this study, we demonstrate the dynamics of a single antiskyrmion confined within an elemental $(Fe_{0.67}Ni_{0.3}Pd_{0.07})_3P$ (FeNiPdP) lamella nanostructure during the application of nanosecond current pulses. Under a fixed magnetic field, we achieve the creation and deletion of a single antiskyrmion by alternating the density of single current pulses. Additionally, we realize a variety of current-controlled topological magnetic transformations among individual antiskyrmions, skyrmions, bubbles, and ferromagnets (FMs). Our findings showcase the versatility of current-controlled antiskyrmion dynamics, which could pave the way for antiskyrmion-based topological spintronic devices.

**Magnetic hysteresis in the nanostructured cuboid**

We grew FeNiPdP bulk single crystals using the self-flux method. Magnetization measurements at 100 mT reveal a Curie temperature of approximately 390 K (Supplementary Fig. 1). Magnetic field-dependent magnetization curves along the *c*-axis and in the *ab*-plane demonstrate perpendicular magnetic anisotropy across the measured temperature range (2 K to 400 K). The uniaxial magnetic anisotropy constant ($K_u$) was calculated from the difference in the areas under the magnetization curves $\int_0^{B_{Sat}} M(B)\mathrm{d}B$ along the *c*-axis and perpendicular to it, up to saturation $B_{Sat}$ (Supplementary Fig. 1).

We fabricated highly confined FeNiPdP nanostructured cuboids with [001] out-of-plane orientation. Electron diffraction and high-resolution transmission electron microscopy (TEM) confirm cuboid edges closely aligned with [110] and [1$\bar{1}$0] crystallographic directions (Supplementary Fig. 2). Lattice parameters *a* and *b* were both determined as 0.92 nm, matching established values.[37] The period ($\lambda$) of zero-field stripe domains in a large FeNiPdP lamella is about 200 nm in the measured temperature range from 95-350 K (Supplementary Fig. 3). Within the rectangle dimension of about 2$\lambda$-3$\lambda$, the confined FeNiPdP cuboid stabilizes a single magnetic soliton. Around and below the critical size (length/width ~1$\lambda$ ≈ 200 nm), edge-confined fractional half-antiskyrmion emerges (Supplementary Fig. 4).[50] Conversely, dimensions > 3$\lambda$ permit multiple solitons, including skyrmion-antiskyrmion pairs under field variation (Supplementary Fig. 5).

Initially, we investigated how magnetic fields influence the evolution of magnetism within a 450-nm FeNiPdP nanostructured cuboid at a small tilted field angle 1.5°, defined as the angle between the external magnetic field and the normal [001] axis of the thin cuboid plane, as shown in Fig. 1. At zero magnetic field, we observe a distorted U-shaped spin texture (Fig. 1b). Because the U-shaped spin texture originates from the deformation of the rectangular antiskyrmion during the field-decreasing process and is topologically equivalent to the antiskyrmion, we call it U-antiskyrmion. As we gradually increase the field, the U-shaped antiskyrmion first contracts and morphs into a rectangular antiskyrmion (Fig. 1a). With further increments in the magnetic field, we obtain multiple topological transformations: the antiskyrmion transforms into a bubble, then a skyrmion, and ultimately into FM. Conversely, during the field-decreasing process, we observe notable hysteresis. Specifically, FM remains stable in fields above 380 mT, transitioning back to an antiskyrmion once the field decreases below 380 mT (Fig. 1b). As the field approaches zero, the rectangular antiskyrmion expands and reverts to a U-shaped antiskyrmion. This field-induced magnetic hysteresis (Fig. 1c) hints at the coexistence of four distinct states—antiskyrmion, skyrmion, bubble, and ferromagnet—at a fixed field. Our simulations accurately reproduce the transition from the U-shaped antiskyrmion at low magnetic fields to the rectangular-shaped antiskyrmion at high magnetic fields (Figs. 1e and 1f). A detailed analysis of the energy terms indicates that the U-shaped antiskyrmion is predominantly stabilized by demagnetization and DMI energies (Supplementary Fig. 6). Moreover, at low magnetic fields, uniaxial anisotropy also contributes to the greater stabilization of the U-shaped antiskyrmion compared to the rectangular-shaped antiskyrmion. The total free energy of the U-shaped antiskyrmion is lower than that of the rectangular-shaped antiskyrmion in the low-magnetic-field regime, explaining the formation of U-shaped antiskyrmions at low magnetic fields.

Typically, such magnetic field-induced topological transformations and magnetic hysteresis are observed across a wide temperature range, from 95 to 340 K (Supplementary Fig. 7). The threshold magnetic fields for topological transformations decrease as the temperature rises (Supplementary Fig. 7). We first show the field-induced magnetic hysteresis. As the field increases from 385 to 510 mT, the initial antiskyrmion transforms into a skyrmion. Moreover, the skyrmion remains stable when the field is decreased back to 385 mT (Supplementary Fig. 8). By utilizing this field-swapping method, we can obtain all four states—antiskyrmion, skyrmion, bubble, and FM—at the same field value (Supplementary Fig. 8). We also explore thermal hysteresis at a fixed field (Supplementary Fig. 9 and Supplementary Fig. 10). We first set antiskyrmion as the initial state at 95 K. In the temperature-increasing process under the application of a fixed field of 417 mT, the antiskyrmion can transform into a bubble at 245 K, and the bubble state remains stable in the temperature-decreasing process. Similarly, the skyrmion and FM states at 95 K and 417 mT can be achieved in a temperature-decreasing process from 250 K and 261 K, respectively. The field- and thermal-induced magnetic hysteresis proves the coexistence of four distinct states at the same field and temperature, enabling the potential controlled transformations between the four states.

We further characterize field- and temperature-dependent antiskyrmion/skyrmion dimensions, defined as lengths $L_x$ and $L_y$ along the [110] and [1$\bar{1}$0] crystallographic axes (Supplementary Fig. 11 and Supplementary Fig. 12). At low fields, antiskyrmions and skyrmions manifest distinct anisotropies $(L_x \neq L_y)$ arising from dipolar interactions and geometrically tunable via nanostructuring. This anisotropy persists for field tilt angles $\alpha \approx 0°–2.5°$, with antiskyrmion long axes consistently aligned to the nanostructure's long axis. Increasing $\alpha$ at a fixed field first transforms antiskyrmions to bubbles while preserving shape anisotropy (Supplementary Fig. 13); further tilt reorients the bubble's long axis toward the in-plane field component to minimize Zeeman energy. Field elevation reduces both $L_x$ and $L_y$, driving convergence toward cubic symmetry at high fields (Supplementary Fig. 11 and Supplementary Fig. 12). Upon increasing the temperature, the transformation field decreases, but the skyrmion/antiskyrmion size follows a similar variation rule in all the measured temperature range. Simulations confirm that nanostructure geometry governs antiskyrmion shape anisotropy (Supplementary Fig. 14).

We then delve into the stability of different magnetic solitons in a strongly confined nanostructure through simulations, at a slight field tilt angle of 2°. By examining magnetic field-dependent evolution and calculating the total free energy, we identify the most stable phases across various field regions (Supplementary Fig. 15). At lower fields, antiskyrmions tend to adopt a distorted U-shaped geometry (Fig. 1d). As the field strengthens, the most stable phase sequentially transitions to a rectangular antiskyrmion, then a bubble, another antiskyrmion, and ultimately the FM state. These field-dependent magnetic profiles (Supplementary Fig. 16) align well with our experimental Fresnel images and retrieve magnetizations of field-driven magnetic evolution (Fig. 1). At low magnetic fields, antiskyrmions are energetically favored over skyrmions due to competing energy contributions (Supplementary Fig. 17). The DMI stabilizes spin twists with anti-vorticity within the domain walls, while exchange energy favors skyrmions because antiskyrmions incur higher exchange costs from Bloch lines. Detailed energy analysis confirms demagnetization and uniaxial anisotropy energies further stabilize antiskyrmions at low fields. Conversely, Zeeman energy preferentially stabilizes skyrmions at low fields. The net energy balance ultimately determines antiskyrmion predominance.

Notably, the field-driven energy evolution can be adjusted by varying the angle of the applied field (Supplementary Fig. 15). A slight tilt is necessary for the four-state transformation. The four-state transformation occurs for field tilt angles of up to approximately 3°. In the absence of a field tilt, the bubble state cannot serve as an intermediate phase during the transformation from antiskyrmion to skyrmion (Supplementary Fig. 15b). Consequently, the field increase yields a direct three-state sequence: antiskyrmion to skyrmion to FM state. Conversely, under a large tilt magnetic field, the skyrmion phase cannot remain the most stable phase across all field regions (Supplementary Fig. 15f), producing a distinct transformation pathway: antiskyrmion to bubble to FM state. The coexistence of different metastable phases within the same field regions accounts for magnetic hysteresis (Fig. 1c). Our experiments confirm the theoretically predicted dependence of magnetic field-driven topological transformations on tilt angle (Supplementary Fig. 18).

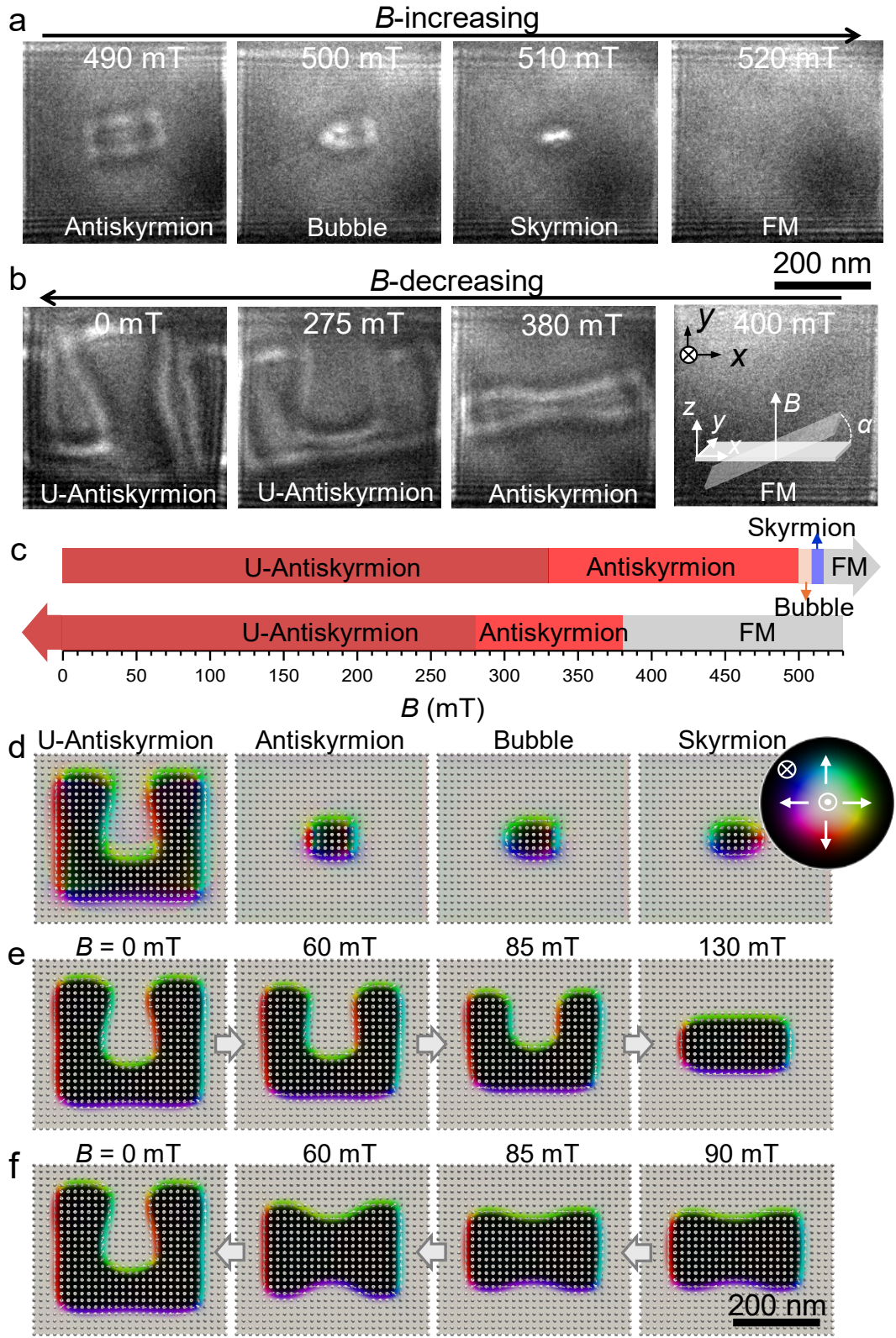

**Fig. 1 | Magnetic hysteresis in the FeNiPdP nanostructured cuboid.** **a** Magnetic evolution in the field-increasing process. **b** Magnetic evolution in the field-decreasing process. Field tilt angle, 1.5°. The Lorentz Fresnel images in (a) and (b) were taken at

a defocus distance of −500 μm. The inset shows a schematic of the experimental field tilt angle, $\alpha$. The in-plane magnetic field component is along the in-plane x-axis. **c** Summary of the stable diagram in the field-increasing and field-decreasing processes. Temperature in (a)-(c), 95 K. **d** Corresponding middle-layer magnetizations of simulated magnetic U-antiskyrmion at 0 mT, antiskyrmion at 290 mT, bubble at 295 mT, and skyrmion at 300 mT. **e** Field-induced morphological evolution starting from a U-shaped antiskyrmion during increasing field. **f** Morphological evolution starting from a rectangular antiskyrmion during decreasing field. The color in (d)-(f) represents magnetization orientation according to the colorwheel.

**Current-controlled dynamics of single antiskyrmions**

We delve deeper into the behavior of antiskyrmions in response to external pulsed electrical stimuli. Starting with the FM state as our initial condition at a constant field of 385 mT with a tilted angle of 1.5°, we then subject the FeNiPdP microdevice to pulsed currents with a duration of 40 ns (Fig. 2a and Supplementary Movie 1). The FM state exhibits no dynamic response until the current density $j$ exceeds $j_{c1}$ (~5.8×10$^{10}$ A/m²). The creation of an antiskyrmion is observed when the current density $j$ surpasses $j_{c1}$. As we further elevate the current density, the antiskyrmion transforms into a bubble at $j > j_{c2}$ (~23.8×10$^{10}$ A/m²), then into a skyrmion at $j > j_{c3}$ (~24.4×10$^{10}$ A/m²), and ultimately reverts to the FM state at $j > j_{c4}$ (~25.5×10$^{10}$ A/m²). We thus designate the symbols $j_{c1}$, $j_{c2}$, $j_{c3}$, and $j_{c4}$ to represent the minimum current densities required for the transformations: FM to antiskyrmion, antiskyrmion to bubble, bubble to skyrmion, and skyrmion back to FM, respectively. Such four-state transformations can also be realized when the current orientation is reversed (Supplementary Fig. 19). The current density dependence of topological transformations and magnetic hysteresis suggests the electrical control of deterministic topological states by altering the current density.

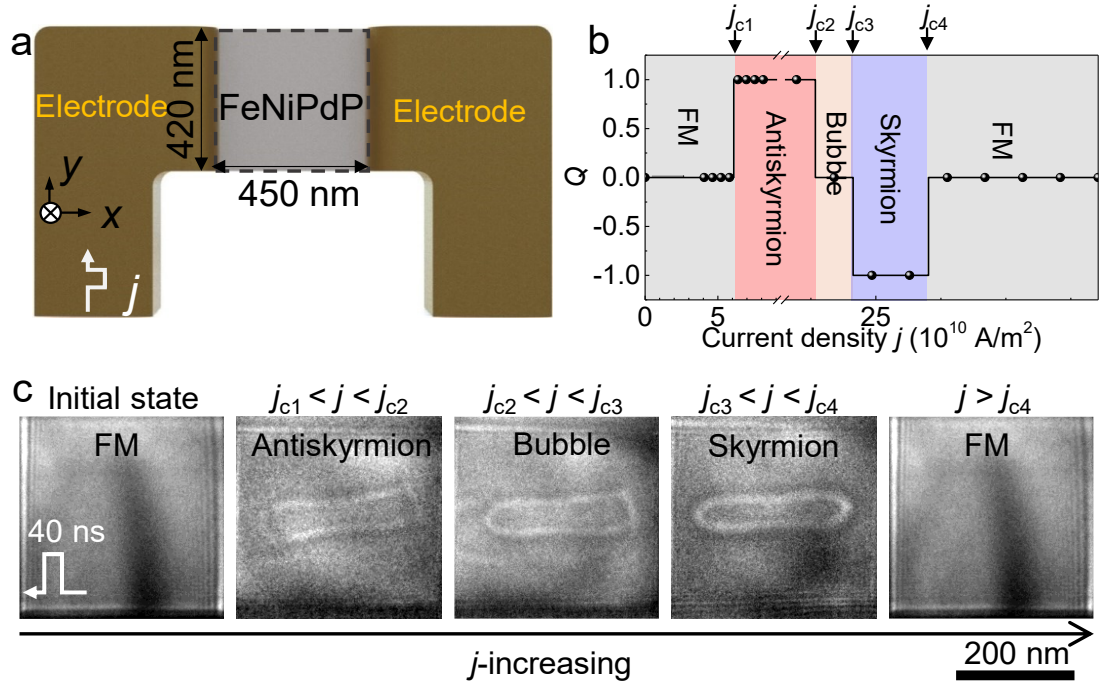

**Fig. 2 | Current density $j$ dependence of topological transformations. a** Schematic profile for the FeNiPdP device. **b** Dependence of topological states on current density starting from $Q = 0$ FM state in the $j$-increasing process. **c** Representative Fresnel imaging of FM, antiskyrmion, bubble, skyrmion, and FM in the $j$-increasing process. Pulse duration 40 ns. Magnetic field, 385 mT. Defocus distance, −500 μm. Field tilt angle, 1.5°. Temperature, 95 K.

*Creating and Deleting Single Antiskyrmions.* We have demonstrated that transitions from FM states to antiskyrmions can be achieved by applying a pulsed current with a density within the range of $j \in [j_{c1}, j_{c2}]$ (Fig. 2). Starting with an antiskyrmion as the initial state, we can achieve an antiskyrmion-to-FM transformation by applying a single current pulse with $j > j_{c4}$. By alternating pulsed currents with densities in the ranges $j \in [j_{c1}, j_{c2}]$ and $j > j_{c4}$, we can reversibly create and delete individual magnetic antiskyrmions (Fig. 3, Supplementary Fig. 8a, and Supplementary Movie 2).

*Skyrmion-Antiskyrmion Transformations.* We have shown that during an increase in current density ($j$-increasing process), an antiskyrmion can transform into a skyrmion through an intermediate bubble state. With an antiskyrmion as the starting point, applying a single current pulse with $j \in [j_{c3}, j_{c4}]$ results in an antiskyrmion-to-skyrmion transformation (Fig. 3). However, directly transitioning from a skyrmion to an antiskyrmion through a single electrical pulse is not feasible. Instead, we achieve the skyrmion-to-antiskyrmion transformation through a two-step process: first, we apply a single current pulse with $j > j_{c4}$ to convert the skyrmion to an FM state, and then we apply another single current pulse with $j \in [j_{c1}, j_{c2}]$ to transform the FM state into an antiskyrmion. In summary, using a double-pulsed current with varying density, we can realize the skyrmion-to-FM-to-antiskyrmion transformation (Supplementary Fig. 20b and Supplementary Movie 3).

*Versatility in Current-Controlled Topological Transformations.* To achieve skyrmion states, we can consistently apply a single current pulse with $j \in [j_{c3}, j_{c4}]$, regardless of whether the initial state is FM, antiskyrmion, or bubble. Similarly, to attain the FM state, we can always use a single current pulse with $j > j_{c4}$, regardless of whether the starting state is antiskyrmion, bubble, or skyrmion. In contrast, achieving the antiskyrmion state necessitates starting from an FM state and applying a single current pulse with $j \in [j_{c1}, j_{c2}]$. The bubble state can only be achieved from an initial FM or antiskyrmion state using a single current pulse with $j \in [j_{c2}, j_{c3}]$. By varying the pulsed current density, we can realize all reversible transformations between four distinct states, including bubble-FM (Supplementary Fig. 20c and Movie 4), skyrmion-FM (Supplementary Fig. 20d and Movie 5), antiskyrmion-bubble (Supplementary Fig. 20e and Movie 6), and skyrmion-bubble (Supplementary Fig. 20f and Movie 7).

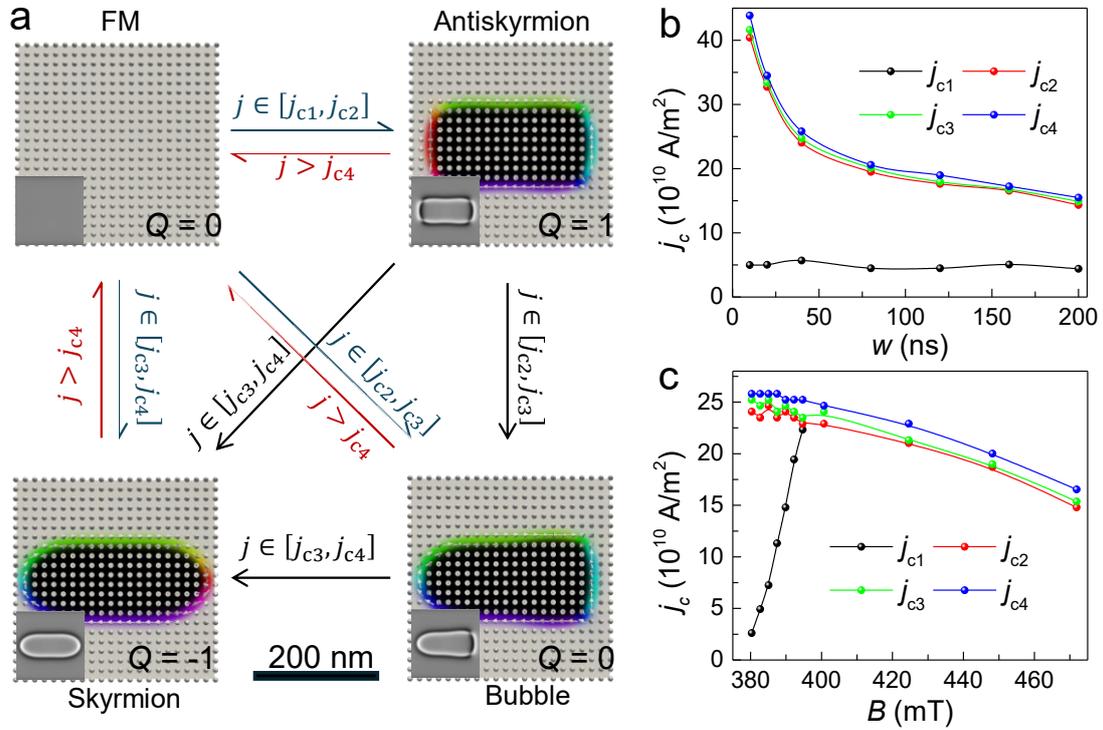

**Fig. 3 | Current-controlled reversible topological transformations occur between four unique phases within the FeNiPdP cuboid. a** Summary for current-controlled topological transformations among single antiskyrmion, skyrmion, bubble, and FM. Insets show the corresponding defocused Fresnel images. A single current pulse with $j > j_{c4}$ deletes antiskyrmions, skyrmions, or bubbles. Writing requires single pulses:

antiskyrmions with $j \in [j_{c1}, j_{c2}]$, bubbles with $j \in [j_{c2}, j_{c3}]$, and skyrmions with $j \in [j_{c3}, j_{c4}]$. Single-pulse transformations include: antiskyrmion to skyrmion ($j \in [j_{c3}, j_{c4}]$), antiskyrmion to bubble ($j \in [j_{c2}, j_{c3}]$), and bubble to skyrmion ($j \in [j_{c3}, j_{c4}]$). Two-step transformations require sequential pulses: skyrmion to antiskyrmion (first $j > j_{c4}$ then $j \in [j_{c1}, j_{c2}]$), bubble to antiskyrmion (first $j > j_{c4}$ then $j \in [j_{c1}, j_{c2}]$), and skyrmion to bubble (first $j > j_{c4}$ then $j \in [j_{c2}, j_{c3}]$). **b** Pulse duration $w$ dependence of threshold transformation current density $j_c$. Magnetic field $B$, 385 mT. **c** Magnetic field $B$ dependence of the threshold transformation current density $j_c$. Pulse duration $w$, 40 ns. Field tilt angle, 1.5°. Temperature, 95 K.

Our voltage source specifications (2-200 ns pulse duration range) permit current-controlled ultrafast topological transformations across this full temporal window (Fig. 3b and Supplementary Fig. 21). As $w$ increases, $j_{c1}$ remains stable around $5.0 \times 10^{10}$ A/m$^2$, while $j_{c2}$, $j_{c3}$, and $j_{c4}$ all decrease (Fig. 3b). We also investigate the dependence of threshold current density ($j_c$) on magnetic field ($B$), as shown in Fig. 3c. As $B$ increases, $j_{c1}$ rises, while $j_{c2}$, $j_{c3}$, and $j_{c4}$ all decline. These differing trends in $j_c$ concerning $w$ and $B$ suggest distinct physical mechanisms underlying the electrically controlled topological transformations. Across a temperature range of 100 to 300 K, the dependence of threshold current densities ($j_{c1}, j_{c2}, j_{c3}$, and $j_{c4}$) on pulse duration $w$ and magnetic field $B$ follow similar patterns (Supplementary Fig. 22). Notably, current-controlled topological transformations among individual antiskyrmions, skyrmions, bubbles, and FM states can be reliably reproduced at room temperature (Supplementary Movie 8).

Furthermore, these current-controlled transformations are highly reproducible in similar devices (Supplementary Fig. 22 and Supplementary Fig. 23). For a large nanostructure with a length × width of 730 nm × 730 nm, a current pulse with a high current density (37.9 × $10^{10}$ A/m²) reliably erases all magnetic solitons (Supplementary Fig. 24). However, these larger nanostructures host diverse magnetic states with varying soliton counts and topologies. Consequently, small current pulses applied to high-energy metastable FM states yield stochastic outcomes, generating 2-4 coexisting solitons (antiskyrmions, skyrmions, and bubbles) with unpredictable configurations.

In accordance with field tilt angle dependence on static field-driven transformations, the current-controlled multi-state transformations are also affected by the tilted field. Our results show distinct operational regimes for a fixed field of 417 mT (Supplementary Fig. 25): below 1.5° enables 3-state transformations (Antiskyrmion ⇌ Skyrmion ⇌ FM); between 1.5° and 3.0° enables 4-state transformations (Antiskyrmion ⇌ Skyrmion ⇌ Bubble ⇌ FM); and above 3.0° enables 3-state transformations (Antiskyrmion ⇌ Bubble ⇌ FM). Therefore, antiskyrmion-skyrmion conversion requires a small field tilt angle, antiskyrmion-bubble conversion requires a large field tilt angle, while antiskyrmion-FM conversion is robust across at all measured angular range. The four-state transformations necessitate a small field tilt, achievable with components like rotatable stages, though this adds complexity to device design. In contrast, three-state transformations (skyrmion–antiskyrmion–FM) do not require field tilting, enabling a more streamlined device architecture.

**Physical origin of the topological magnetic transformations**

We finally explore microscopic mechanisms governing current-induced topological transformations among individual antiskyrmions, skyrmions, bubbles, and FM states. In low-field regions, the rectangular antiskyrmion exhibits the lowest energy, while the FM state has the highest. Starting with the FM state, we investigate its dynamic response to Zhang-Li spin-transfer torque (STT) by applying an in-plane current[51], as shown in Fig. 4a and 4c, and Supplementary Movie 9. Our findings indicate that a nucleation seed forms during the excitation of a spin-polarized current. This seed rapidly expands to create a closed domain with intricate domain walls featuring multiple Bloch points. These complex domain walls do not represent the lowest energy state and eventually transform into an antiskyrmion with $Q = 1$. The oscillation of topological charge observed between 1.4 and 2.4 ns (Fig. 4c) results from annihilation of additional high-energy Bloch points within the initially complex domain walls. Our simulations reveal that the FM-to-antiskyrmion transformation occurs on a very short timescale. For current pulses >2 ns, simulations show pulse-duration-independent threshold current densities for FM-to-antiskyrmion transformation (Supplementary Fig. 26), consistent with our experimental data (Fig. 3c). While prior studies demonstrated ultrafast skyrmion generation using sub-ns pulses,[11] we find that sub-ns antiskyrmion creation via STT demands higher current densities (Supplementary Fig. 26). Furthermore, as the magnetic field increases, our simulations show an increase in $j_{c1}$ (Supplementary Fig. 27), which is in line with experimental results (Fig. 3c).

In contrast, we attribute the transformation from antiskyrmion to bubble to skyrmion to FM state to the combined effects of Joule heating and STT. Our experiment has shown magnetic hysteresis during the temperature-varying process at a constant magnetic field (Supplementary Fig. 8 and Supplementary Fig. 9). Applying current pulses also results in a temperature increase due to Joule heating. The larger the current density, the greater the temperature rise[52]. Once the nanosecond current pulse is turned off, the temperature returns to normal. This process aligns well with the static magnetic evolution observed during temperature changes. Consequently, during ns-duration current pulses, Joule heating contributes significantly to the observed reduction in magnetic soliton size. Upon current termination, both the temperature and soliton size recover. The critical current densities $j_{c2}$, $j_{c3}$, and $j_{c4}$ all decrease as pulse duration $w$ increases (Fig. 3b), which is in line with the typical Joule heating effect. Furthermore, the thermally induced transformation must follow the sequence of antiskyrmion, bubble, skyrmion, and finally FM. States later in the sequence are more stable in high-field regions at a fixed temperature or in high-temperature regions at a fixed field. Here, we establish a current-induced temperature increase based on the experimental temperature dependence of magnetic domain evolution (Supplementary Fig. 9). When the temperature rises from 95 K to 245 K at a fixed magnetic field, the antiskyrmion transforms into a bubble. Applying a 40-ns current pulse with a critical current density of about $2.4 \times 10^{11}$ A/m$^2$ also induces the antiskyrmion-to-bubble transformation. This equivalence implies that such a current pulse induces a temperature rise of approximately 150 K. Furthermore, using the experimental device geometry and material parameters, we simulated the transient temperature variation during pulsed current injection via multiphysics field simulations (Supplementary Fig. 28).[53] Our simulations confirm that a 40-ns pulse at this critical current density generates an average temperature rise of about 200 K, comparable with our experimental estimate (~150 K). Critically, STT efficiency scales with current density, enabling sequential antiskyrmion→bubble→skyrmion→FM transformations through synergistic Joule heating and STT effects. During current pulses, Joule heating dynamically reshapes the magnetic energy landscape, shifting the energy minimum from antiskyrmion to bubble to skyrmion to FM during thermal warming at fixed field (Supplementary Fig. 29). STT subsequently drives transitions from metastable states to the new ground state. Micromagnetic simulations confirm reversibility upon energy landscape modification (Fig. 4d), with field-increase behavior at fixed temperature (Supplementary Fig. 15) mirroring thermal evolution at a fixed field (Supplementary Fig. 29): fields of 310 mT stabilize the FM state, permitting STT-driven antiskyrmion-to-FM transformation. Thus, antiskyrmion→bubble→skyrmion→FM transformations could arise from the two-step mechanism comprising: (1) Joule-heating-induced energy landscape modification and (2) STT-mediated transitions toward equilibrium states.

A low-energy state at high temperatures can persist as a metastable high-energy state even after temperature recovery due to magnetic hysteresis. However, thermal heating effects do not support the reverse transformation from FM to skyrmion, bubble, and then antiskyrmion. Thanks to another mechanism-dominated transformation from FM to antiskyrmion (Fig. 4), we can achieve the reverse transformation from FM to skyrmion or bubble, with the antiskyrmion state serving as an intermediate. The Joule thermal heating is determined only by the density and is not dependent on current orientation. Besides, STT acts as an external stimulus to drive the transformation from the metastable phase to the most stable phase, which is shown not to be dependent on current orientation (Supplementary Fig. 26). Thus, the four-state mutual transformations are not strongly dependent on the current orientation (Supplementary Fig. 19).

Importantly, the definitions of antiskyrmion, skyrmion, and bubble apply exclusively to the middle layers of the FeNiPdP structure. To minimize magnetic dipole-dipole interactions, surface layers consistently adopt Néel-twisted skyrmion configurations with $Q = -1$ (Supplementary Fig. 30).[35,41] Consequently, mutual transformations between antiskyrmion, skyrmion, and bubble states alter the topological charges only in the middle layers, while the surface layers retain their skyrmionic topology. We therefore present the representative 2D magnetization of the middle layer (Figs. 1, 3, 4), which captures the essential topological transformations. The simulated and observed dynamics of these 3D inhomogeneous states under in-plane current are governed by the interplay of Zhang-Li STT and the system's free energy landscape, reflecting an ensemble-averaged response. For instance, a hybrid antiskyrmion tube with surface Néel-type skyrmions can exhibit a Hall balance, as previously identified.[35]

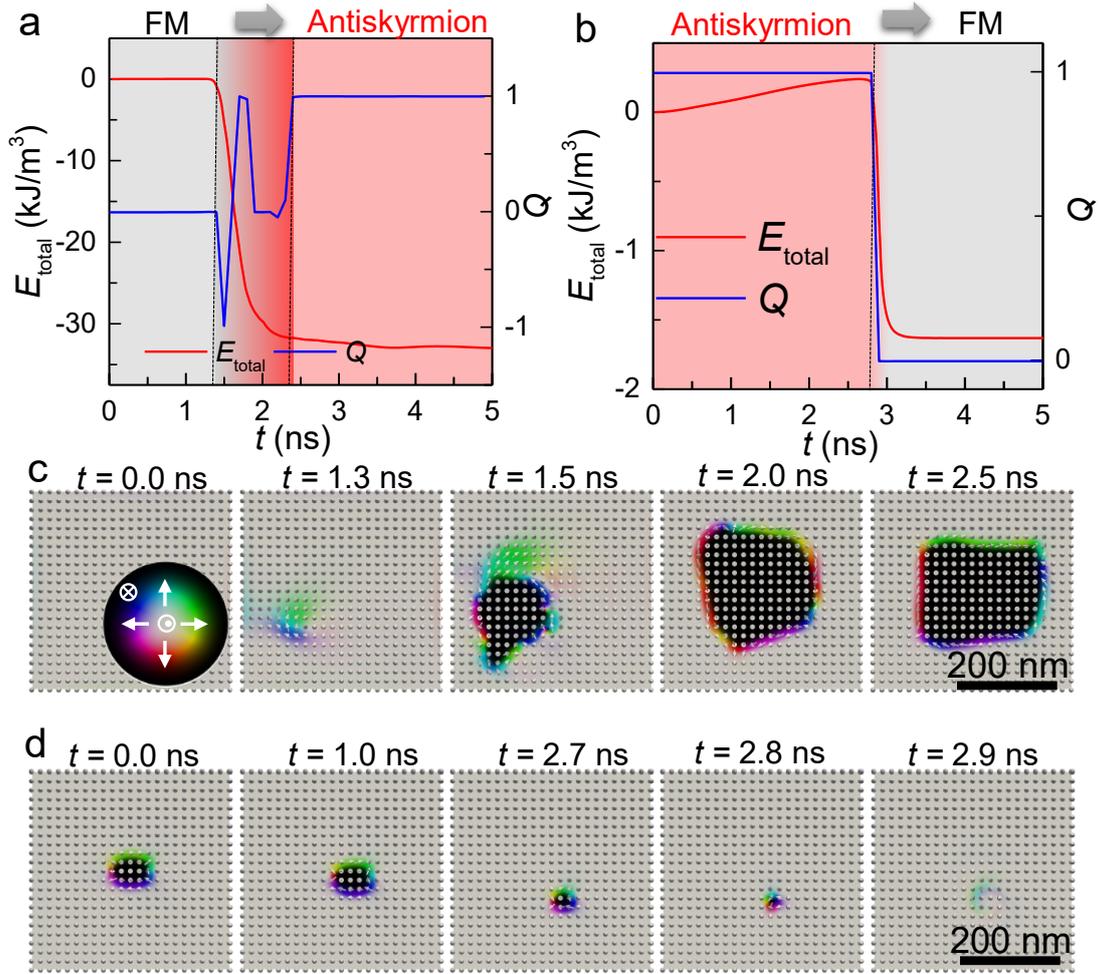

**Fig. 4 | Simulated STT-induced FM-antiskyrmion transformations. a-b** Time $t$ dependence of total free energy and topological charge $Q$ during the FM-antiskyrmion transformation by applying an in-plane current with $j = 5.0 \times 10^{11}$ A/m$^2$. (a) FM-to-antiskyrmion. Magnetic field $B$, 85 mT. (b) Antiskyrmion-to-FM. Magnetic field $B$, 85 mT. **c-d** Representative middle-layer magnetic configurations during the FM-antiskyrmion transformation. (c) FM-to-antiskyrmion. (d) Antiskyrmion-to-FM. Field tilt angle, 2°.

## Discussion

In summary, for the first time, we fulfill the fundamental electrical operations on individual antiskyrmions, including reversible creation, deletion, and topological transformations with skyrmions and bubbles in a nanostructured element. We have established the stabilization mechanisms and mutual topological transformations of individual magnetic solitons in the strongly confined elemental nanostructure as a function of temperature and magnetic field. These current-controlled topological transformations exhibit great versatility and can be achieved over a broad temperature range, including room temperature. The deterministic electrical control of magnetic solitons is explained by current-induced spin-transfer torque, Joule thermal heating, and magnetic hysteresis. Our findings shed light on the potential applications of antiskyrmion-hosting materials, such as in multi-state information processing and memory[34,40].

**Data availability**

The data that support the plots provided in this paper and other findings of this study are available from the corresponding author upon reasonable request.

**Methods**

**Bulk sample preparations**

Single FeNiPdP crystals with $S_4$ symmetry were grown by the self-flux method with stoichiometric iron (Alfa Aesar, >99.9%) and Nickel (Alfa Aesar, >99.9%), palladium (Alfa Aesar, >99.9%), and red phosphorus (Alfa Aesar, >99.9%). The sintered bulk FeNiPdP was obtained by heating the mixture at 1100 °C for 4 days. The crystal orientations were further identified using electron diffraction and high-resolution transmission electron microscopy (Supplementary Fig. 2).

**Fabrication of FeNiPdP microdevices**

We fabricated thin FeNiPdP microdevices from bulk single crystals using a standard lift-off method in a focused ion beam and scanning electron microscopy (FIB-SEM) dual-beam system (Helios Nanolab 600i, FEI). The system was equipped with a gas injection system and a micromanipulator (Omniprobe 200+). We first carved FeNiPdP nanostructure on the surface of the single crystal after milling two trenches. The FeNiPdP nanostructure was then lifted off and transferred close to an electrical chip using the micromanipulator. Each microdevice features two electrical contacts and consists of an FeNiPdP cuboid encased in ion-deposited carbon layers on its top surfaces. The left and right ends of the cuboid were connected to Au electrodes on the chip by fabricating Pt nanosticks. The length and width of the FeNiPdP cuboid are designed to be about 300-700 nm. The FeNiPdP cuboids were all finally thinned to a thickness of approximately 140-150 nm. A detailed fabrication process of microdevices has been presented in a previous study.[19] All main-text data derive from a single 450 nm × 420 nm microdevice. Supplementary figures include full-dimensional documentation for other devices.

**TEM measurements**

We used in-situ Lorentz-TEM (Talos F200X, FEI) to investigate the current-induced magnetic domain dynamics in the FeNiPdP nanostructure.[39,54] The current pulses with pulse durations varying from 2-200 ns were provided by a voltage source (AVR-E3-B-PN-AC22, Avtech Electrosystems). The temperature can be adjusted in the range from 95 to 350 K.

**Micromagnetic simulations**

*Micromagnetic simulations*: The zero-temperature micromagnetic simulations were performed using MuMax3.[55] We consider the Hamiltonian exchange interaction energy ($\varepsilon_{ex}$), 2D chiral DMI energy ($\varepsilon_{D2d}$), uniaxial magnetic anisotropy energy ($\varepsilon_u$), Zeeman energy ($\varepsilon_{zeeman}$), and magnetic dipole-dipole interaction energy ($\varepsilon_{dem}$) [55]. The total energy terms are given by:

$$\varepsilon = \int_{V_s} \{\varepsilon_{ex} + \varepsilon_u + \varepsilon_{zeeman} + \varepsilon_{dem} + \varepsilon_{D2d}\} d\boldsymbol{r} \tag{1}$$

Here, exchange energy $\varepsilon_{ex} = A(\partial_x \mathbf{m}^2 + \partial_y \mathbf{m}^2 + \partial_z \mathbf{m}^2)$, $\varepsilon_{D2d} = D_{2d} \mathbf{m} \cdot (y \partial_y \mathbf{m} - x \partial_x \mathbf{m})$, uniaxial magnetic anisotropy energy $\varepsilon_u = -K_u (\mathbf{m} \cdot \mathbf{n}_u)^2$, Zeeman energy $\varepsilon_{zeeman} = -M_s \mathbf{B}_{ext} \cdot \mathbf{m}$, and demagnetization energy $\varepsilon_{dem} = -\frac{1}{2} M_s \mathbf{B}_d \cdot \mathbf{m}$. Here, $\mathbf{m} \equiv \mathbf{m}(x, y, z)$ is the normalized units continuous vector field that represents the magnetization $\mathbf{M} \equiv M_s \mathbf{m}(x, y, z)$. The parameters $A$, $D_{2d}$, $K_u$, and $M_s$ are the exchange interaction, DMI, uniaxial anisotropy constant, and saturation magnetization, respectively. $\mathbf{n}_u$ is the unit vector field of the uniaxial easy magnetization axis and is set along the (001) z-axis. $\mathbf{B}_d$ is the demagnetization field. In the absence of the $\varepsilon_{D2d}$ energy term, the system's total energy landscape mirrors that of conventional bubble materials in uniaxial magnets and can stabilize both type-I and type-II bubbles.

We set the exchange stiffness as $A$ = 8.0 pJ/m and chiral interaction as $D_{2d}$ = 0.25 mJ/m$^3$. We set $K_u$ = 110 kJ/m$^3$ and $M_s$ = 617 kA/m based on measured parameters extracted from macroscopic magnetization measurements (Supplementary Fig. 1). Our previous study reveals that the experimental spin textures in FeNiPdP can be well reproduced by simulation based on these magnetic parameters.[35] These magnetic parameters are close to those widely used in previous studies on similar FeNiPdP materials.[31,37] Simulation considering thermal fluctuation field indicated a Curie temperature of 420 K (Supplementary Fig. 31), which is close to that of experiments. The cell size was set at $2 \times 2 \times 2$ nm$^3$. A Zhang-Li spin-transfer torque (STT) is considered for simulating current-driven dynamics. We set the Gilbert damping of 0.05 and the non-adiabatic parameter of 0.

**Acknowledgments**

This work was supported by the National Key R&D Program of China, Grant No. 2024YFA1611303 (J.T.); the Natural Science Foundation of China, Grants No. 52325105 (H.D.), 12422403 (J.T.), 12174396 (J.T.), 12104123 (Y.W.), U24A6001 (J.T.), and 12241406 (H.D.); the Anhui Provincial Natural Science Foundation, Grants No. 2308085Y32 (J.T.) and 2508085MA018 (Y.W.); the Natural Science Project of Colleges and Universities in Anhui Province, Grants No. 2022AH030011 (J.T.) and 2024AH030046 (Y.W.); Anhui Province Excellent Young Teacher Training Project, Grant No. YQZD2023067 (Y.W.); the 2024 Project of GDRCYY (No. 217, Yaodong Wu); the China Postdoctoral Science Foundation Grant No. 2023M743543 (Y.W.); CAS Project for Young Scientists in Basic Research, Grant No. YSBR-084 (H.D.); and Systematic Fundamental Research Program Leveraging Major Scientific and Technological Infrastructure, Chinese Academy of Sciences, Grant No. JZHKYPT-2021-08 (H.D.). We thank the staff members of the Physical Property Measurement System (https://cstr.cn/31125.02.SHMFFPPMS) at the Steady High Magnetic Field Facility, CAS (https://cstr.cn/31125.02.SHMFF), for providing technical support and assistance in data collection and analysis.

**Author contributions**

H.D. and J.T. supervised the project, conceived the idea, and designed the experiments. W.L. synthesized the FeNiPdP bulk crystals. Y.W. fabricated the microdevices and performed the TEM measurements with the help of J.J. and J.T.. J.T. performed the simulations with the help of L.K. and H.Z.. J.T., Y.W., J.J., and H.D. wrote the manuscript with input from all authors. M.T. and S.W. discussed the results and contributed to the manuscript.

**Competing interests**

The authors declare no competing interests.

**Additional information**

Supplementary information is available for this paper.